\documentclass[runningheads]{llncs}
\usepackage{graphicx}
\usepackage{times}
\usepackage{epsfig}
\usepackage{graphicx}
\usepackage{amsmath}
\usepackage{amssymb}
\usepackage{amsfonts}
\usepackage{bm}
\usepackage{subfig}
\usepackage{multirow}
\usepackage{makecell}
\usepackage{pdfx}
\hypersetup{pagebackref=true,breaklinks=true,colorlinks,bookmarks=false}
\newcommand{\myparatight}[1]{\smallskip\noindent{\bf {#1}:}~}

\begin{document}
\title{Understanding the Security of Deepfake Detection}
\author{Xiaoyu Cao, Neil Zhenqiang Gong}
 \institute{Duke University, Durham, NC, USA\\
\email{\{xiaoyu.cao, neil.gong\}@duke.edu}}
\maketitle              

\begin{abstract}
Deepfakes pose growing challenges to the trust of information on the Internet.
Thus, detecting deepfakes has attracted increasing attentions from both academia and industry. 
State-of-the-art deepfake detection methods consist of two key components, i.e., \emph{face extractor} and \emph{face classifier}, which extract the face region in an image and classify it to be real/fake, respectively.  
Existing studies mainly focused on improving the detection performance in \emph{non-adversarial settings}, leaving 
  security of  deepfake detection in \emph{adversarial settings} largely unexplored. 
In this work, we aim to bridge the gap. In particular, we perform a systematic measurement study to understand the security of the state-of-the-art deepfake detection methods in adversarial settings. We use two large-scale public  deepfakes data sources including FaceForensics++ and Facebook Deepfake Detection Challenge, where the deepfakes are fake face images; and we train state-of-the-art deepfake detection methods. These detection methods can achieve  0.94--0.99 accuracies in non-adversarial settings on these datasets. However, our measurement results uncover multiple security limitations of the deepfake detection methods in adversarial settings. 
First, we find that an attacker can evade a face extractor, i.e., the face extractor fails to extract the correct face regions, via adding small Gaussian noise to its deepfake images.  Second, we find that a face classifier trained using deepfakes generated by one method cannot detect deepfakes generated by another method, i.e., an attacker can evade detection via generating   deepfakes using a new method. Third, we find that an attacker can leverage \emph{backdoor attacks} developed by the adversarial machine learning community to evade a face classifier. Our results highlight that deepfake detection should consider the adversarial nature of the problem.


\end{abstract}
\section{Introduction}

Deepfakes generally refer to forged media such as images and videos. 
While forged media has been in existence for decades and was conventionally created by computer graphics methods~\cite{faceswap,thies2016face2face}, recent progress in deep learning  enables automatic,  large-scale creation of realistic-looking deepfakes. In particular, many methods (e.g., generative adversarial networks \cite{karras2017progressive,karras2019style,choi2018stargan,park2019semantic,karras2020analyzing}) have been proposed to generate deepfakes, which we call \emph{deepfake generation methods}. Such deepfakes can further be widely propagated on social medias to spread propaganda, disinformation, and fake news. For instance, comedian Jordan Peele produced a fake video of President Obama criticizing President Trump by altering the lip movements of Obama in a real video~\cite{obamafake}. As a result, deepfakes introduce grand challenges to the trust of online information. In this work, we focus on fake faces, a major category of deepfakes, because faces are key components in human communications and their forgeries lead to severe consequences. Therefore, we will use deepfakes and fake faces interchangeably throughout the paper.

\begin{figure}[!t]
    \center
    \includegraphics[width=0.99\textwidth]{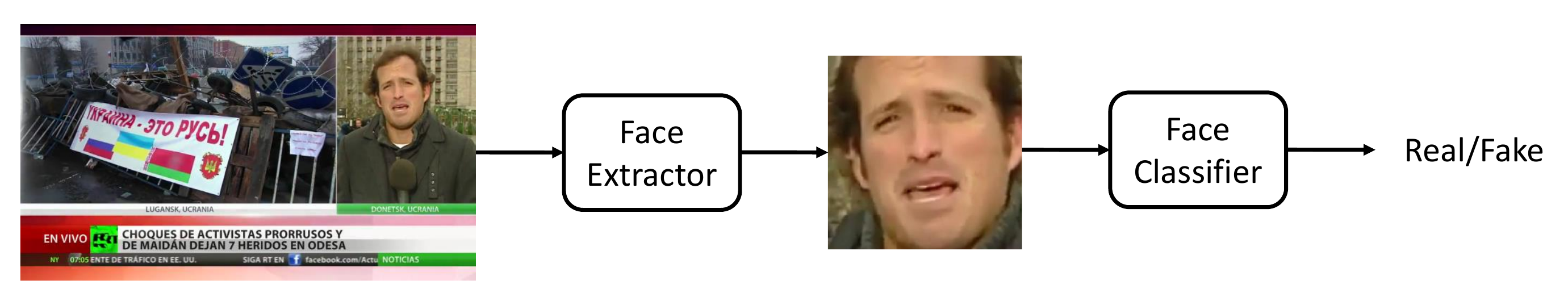}
    \caption{Illustration of the two key components of a deepfake detection system.}
    \label{overview}
\end{figure}

Due to the growing concerns of deepfakes, detecting deepfakes has attracted increasing attentions from both academia and industry. For instance, 
Facebook recently launched a deepfake detection competition~\cite{DFDC2020} to facilitate the development of deepfake detection methods. 
A deepfake detection system includes two key components, i.e., \emph{face extractor} and \emph{face classifier}, which is illustrated in Figure~\ref{overview}. Specifically, a face extractor extracts the face region in an image, while a face classifier classifies the extracted face region to be real or fake. As face extraction is a mature technique, existing deepfake detection methods often use state-of-the-art face extractor but adopt different face classifiers. 
Roughly speaking, existing face classifiers can be grouped into two categories, i.e., \emph{heuristic-based} and \emph{neural-network-based}. 
Heuristic-based face classifiers  \cite{agarwal2019protecting,li2018ictu,matern2019exploiting,yang2019exposing,li2019exposing,frank2020leveraging} rely on some heuristics to distinguish between fake faces and real faces. For instance, Li et al. \cite{li2018ictu} designed a face classifier based on the observation  that  the eyes in fake  faces did not blink normally as people do in the real world. 
However, these heuristic-based face classifiers were soon broken by new fake faces. For instance, the eye-blinking based face classifier was easily broken by fake faces that blink eyes normally~\cite{agarwal2019protecting}.

Neural-network-based face classifiers \cite{zhou2017two,afchar2018mesonet,roessler2019faceforensicspp,nguyen2019multi,nguyen2019use,cozzolino2017recasting,bayar2016deep,rahmouni2017distinguishing,wang2019cnngenerated} 
train deep neural networks to distinguish between fake faces and real faces. Specifically, given a training dataset consisting of both real faces and fake faces generated by some deepfake generation methods, a deep neural network classifier is trained. Then, given a new face, the deep neural network outputs a label that is either ``real'' or ``fake'' for it. These deep neural network based face classifiers achieve the state-of-the-art detection accuracy, showing promising applications in detecting fake faces. Existing studies mainly focused on improving the detection performance under \emph{non-adversarial settings}, i.e., they assume the attacker who generates the deepfakes does not adapt to the detectors. Deepfake detection is essentially a security problem, in which an attacker always adapts to defenses. However, the security of the state-of-the-art deepfake detection methods in such adversarial settings remains largely unexplored, except that a few studies~\cite{gandhi2020adversarial,carlini2020evading,hussain2020adversarial,fernandes2020adversarial} showed that face classifiers are vulnerable to \emph{adversarial examples}~\cite{szegedy2013intriguing,carlini2017towards}. In particular,  an attacker can add carefully crafted perturbations to its fake faces such that a face classifier is highly likely to misclassify them as real faces.

\myparatight{Our work} In this work, we perform systematic measurement studies to understand the security of deepfake detection. The security of a deepfake detection method relies on the security of both the face extractor and the face classifier. Therefore, we perform measurement studies to understand the security of both components. 
In particular, we aim to study the following three questions, where Q1 is related to the security of face extractor while Q2 and Q3 are related to the security of face classifier:
\begin{itemize}
    \item {\bf Q1:} An attacker can easily add perturbations (e.g., Gaussian noise) to its deepfake images. Therefore, the first question we aim to study is: can a face extractor still successfully extract the face region in a deepfake image when an attacker adds a small perturbation to it?  
    \item {\bf Q2:}  Existing studies~\cite{afchar2018mesonet,roessler2019faceforensicspp} often train and test a face classifier using fake faces that are generated by the same deepfake generation methods. However, many different methods have been proposed to generate deepfakes and new methods are continuously developed. 
    Therefore, the second question we aim to study is: can a face classifier trained using fake faces generated by some deepfake generation methods correctly classify fake faces generated by a different method?
    \item {\bf Q3:} A face classifier is a binary machine learning classifier. The adversarial machine learning community has developed \emph{adversarial examples} and \emph{data poisoning attacks} for machine learning classifiers, which attack classifiers at testing phase and training phase, respectively. Recent studies showed that a face classifier is  vulnerable to adversarial examples. The third question we aim to study is: are face classifiers also vulnerable to  data poisoning attacks?

\end{itemize}

{\bf Measurement setup.} 
We extract six deepfakes datasets from two large-scale public sources, i.e., FaceForensics++~\cite{roessler2019faceforensicspp} and Facebook Deepfake Detection Challenge (DFDC) \cite{DFDC2020}. We use Dlib~\cite{dlib09}, an open source face extractor, to extract faces from the images in the datasets. These datasets include 0.9 million to 1.3 million real or fake faces. 
Among the four datasets extracted from FaceForensics++, each of them contains real faces and their corresponding fake faces that are generated by a specific deepfake generation method.  DFDC contains fake faces that are generated by 8 deepfake generation methods. We divide each dataset into three partitions, i.e., \emph{training}, \emph{validation}, and \emph{testing}. Moreover, we train a state-of-the-art neural network based face classifier for each dataset using its training and validation partitions, where the validation partition is used to select the face classifier with the best performance during training. These face classifiers are very accurate in non-adversarial settings, i.e., they achieve  0.94--0.99 accuracies on the testing faces. 

{\bf Measurement results.}
For {Q1}, we add different amount of random Gaussian noise to an image. Then, we use Dlib to extract the face region from a noisy image.  We find that a small Gaussian noise can spoof the face extractor. For instance, when adding noise sampled from a Gaussian distribution with zero mean and standard deviation $\sigma=0.3$ to images in the DFDC dataset, Dlib fails to extract the face regions for at least 80\% of the images. Our results show that face extractor is not secure against small random noise added to an image.

For Q2, we measure the \emph{cross-method generalization} of the face classifiers.  Specifically, we assume the face regions are successfully extracted.  We use a face classifier trained on one dataset to classify the testing faces in another dataset, where the fake faces in the two datasets are generated by different deepfake generation methods. We find that the face classifiers' accuracies drop to nearly random guessing (i.e., 0.5) in such cross-method setting. Our results show that a face classifier trained on deepfakes generated by one method cannot correctly classify deepfakes generated by another method. Our results imply that the standard evaluation paradigm in previous work~\cite{afchar2018mesonet,roessler2019faceforensicspp}, which trains and tests a face classifier using fake faces generated by the same deepfake generation methods, is insufficient to characterize the security of the face classifier.

For Q3, we measure the security of a face classifier against one simple data poisoning attack called \emph{label flipping attack} and one advanced data poisoning attack called \emph{backdoor attack}~\cite{Gu17}. Like Q2, we assume the face regions have been successfully extracted and focus on the security of face classifier. 
Label flipping attack changes the labels of some training faces such that a trained face classifier is corrupted. We find that the face classifiers are relatively secure against the simple label flipping attacks. For instance, their accuracies only drop by 0.04 when the labels of around 25\% of their training faces are flipped. 
However, the more advanced backdoor attacks can break the face classifiers. In a backdoor attack, an attacker adds a \emph{trigger} (e.g., a chessboard grid in our experiments) to some training faces (e.g., we use 5\% of training faces) and changes their labels to be real. Then, a face classifier is trained on the poisoned training faces. We find that the face classifier misclassifies a fake face as real once we inject the same trigger to the fake face.

{\bf Security implications.} Our measurement results have several implications for the security of  deepfake detection. In particular, our results show that an attacker can evade detection of its deepfakes by adding small Gaussian noise  to them to spoof face extraction; an attacker can evade a face classifier by generating deepfakes using a new deepfake generation method; and an attacker can  evade detection by exploiting backdoor attacks to face classifiers.  Our results highlight that deepfake detection should consider the adversarial nature of the problem and take strategic adaptive attacks into consideration.

We summarize our key contributions as follows:
\begin{itemize}
    \item We perform the first systematic measurement study to understand the security of the state-of-the-art deepfake detection methods in adversarial settings.
    \item We find that face extractor is not secure against small perturbation added to an image.
    \item We find that face classifier is not secure against new deepfake generation methods and backdoor attacks.
\end{itemize}

\begin{figure*}[!t]
    \scalebox{1}{\includegraphics[width=0.99\textwidth]{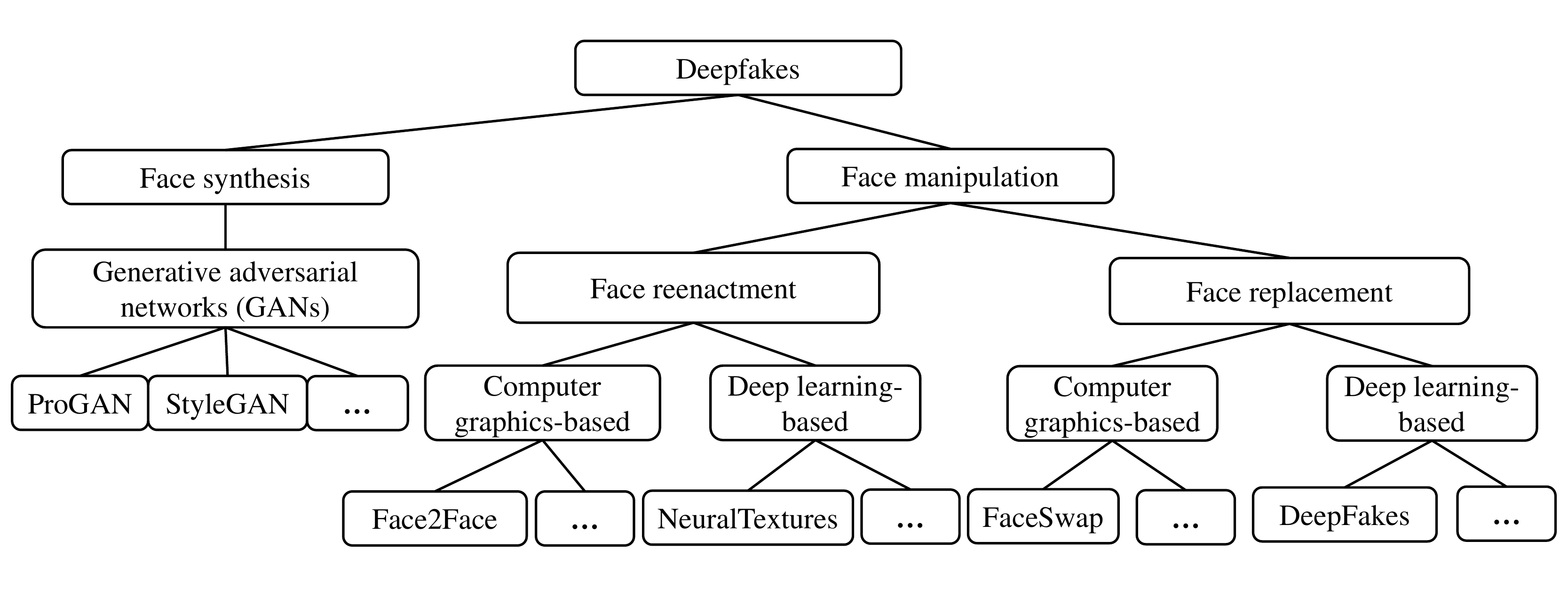}}
    \caption{A taxonomy of deepfakes and their generation methods.}
    \label{fig:df_tree}
\end{figure*}

\section{Background and Related Work}

Deepfakes, a combination of ``deep learning'' and ''fake''~\cite{dfwiki}, usually refer to media such as images and videos that are forged by deep learning methods. In this work, 
we consider deepfakes as  faces forged by both deep learning methods and conventional computer graphics methods. 
Figure \ref{fig:df_tree} shows a taxonomy of deepfakes and  their generation methods. Deepfakes for faces roughly include \emph{face synthesis} and \emph{face manipulation}. Face synthesis aims to synthesize faces that do not belong to any persons in the real world, while face manipulation aims to tamper a person's face image to change its facial expression or completely replace it as another person's face. Next, we discuss face synthesis and face manipulation separately.

\subsection{Face Synthesis}
Most face synthesis methods \cite{karras2017progressive,karras2019style,choi2018stargan,park2019semantic,karras2020analyzing,zhu2017unpaired} leverage the popular deep learning methods called \emph{generative adversarial networks (GANs)}  \cite{goodfellow2014generative}. 
A GAN has two main components, i.e., a \emph{generator}  and a \emph{discriminator}. Both generator and discriminator are neural networks.  The generator takes a random vector (e.g., a vector of random Gaussian noise) as input and  aims to generate a fake face that cannot be distinguished from real faces by the discriminator, while the discriminator aims to distinguish between  real faces and the fake faces generated by the generator. The two neural networks are trained in turn until convergence, after which the generator wins the race and is able to generate fake faces that cannot be distinguished from real faces by the discriminator. Many GANs have been proposed to synthesize faces. Examples include ProGAN \cite{karras2017progressive}, StyleGAN \cite{karras2019style}, StyleGAN2 \cite{karras2020analyzing}, StarGAN \cite{choi2018stargan}, and CycleGAN \cite{zhu2017unpaired}, etc.. For instance,  StyleGAN can synthesize fake faces with given styles. StyleGAN2 further improves the fake face quality by redesigning the generator of the StyleGAN.  The fake faces synthesized by StyleGAN2 are illustrated on a website \cite{thispersondoesnotexist},   refreshing which shows a new synthesized fake face each time.

\subsection{Face Manipulation}

\begin{figure}[!t]
    \center
    \subfloat[Real]{\includegraphics[width=0.3\textwidth]{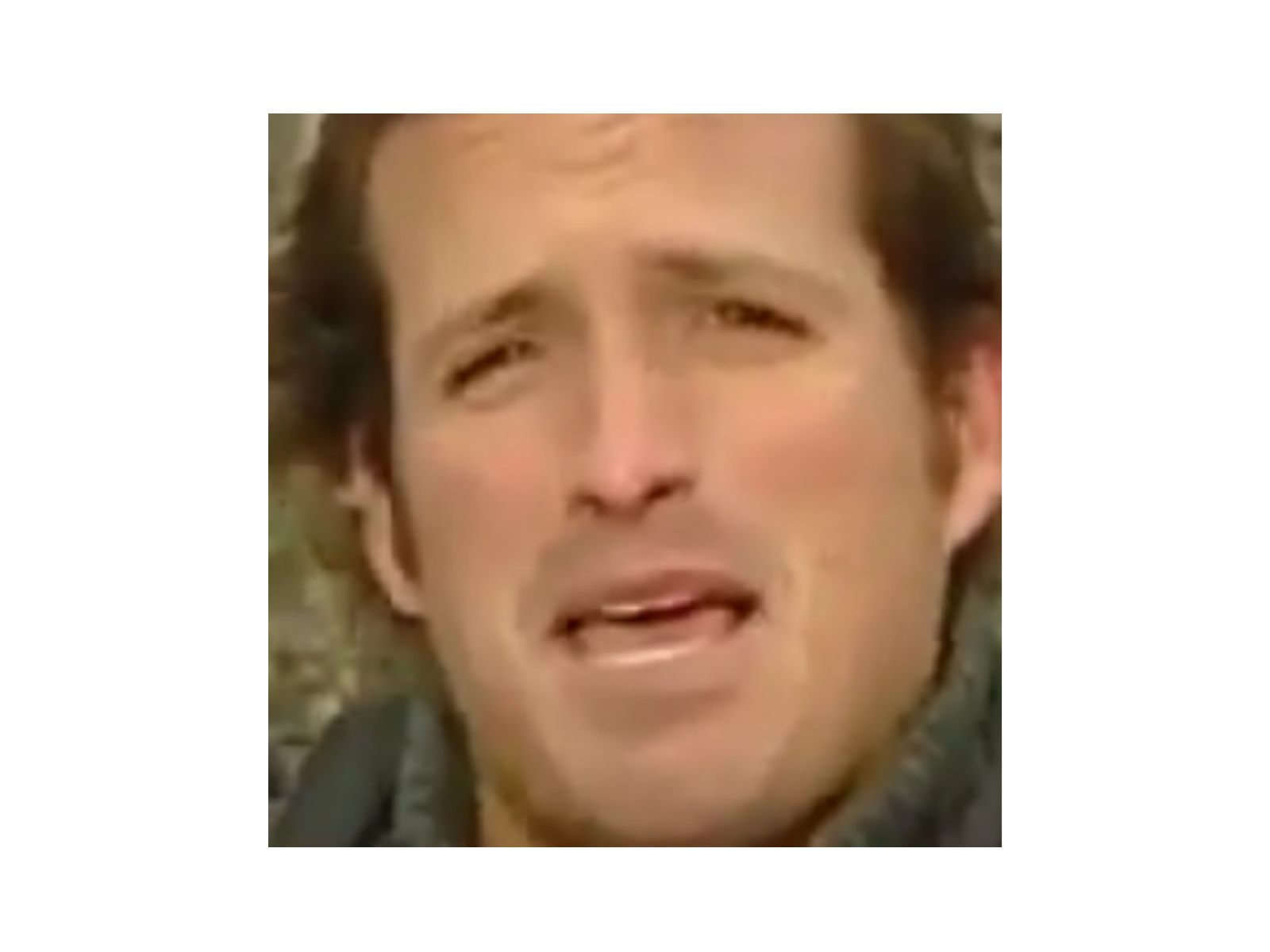}\label{fig:real_example}}  
    \subfloat[Face reenactment]{\includegraphics[width=0.3\textwidth]{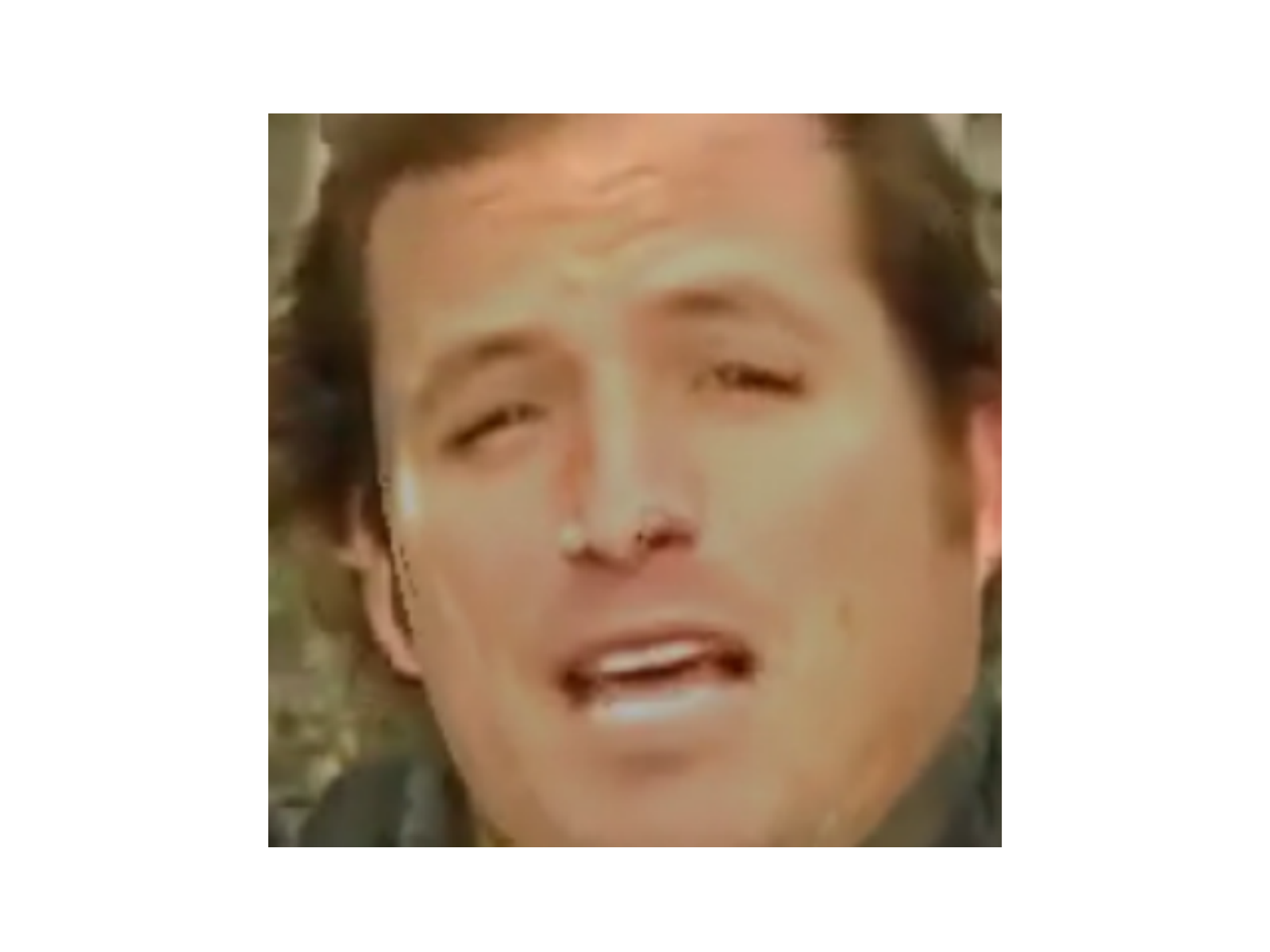}\label{fig:f2f_example}}
    \subfloat[Face replacement]{\includegraphics[width=0.3\textwidth]{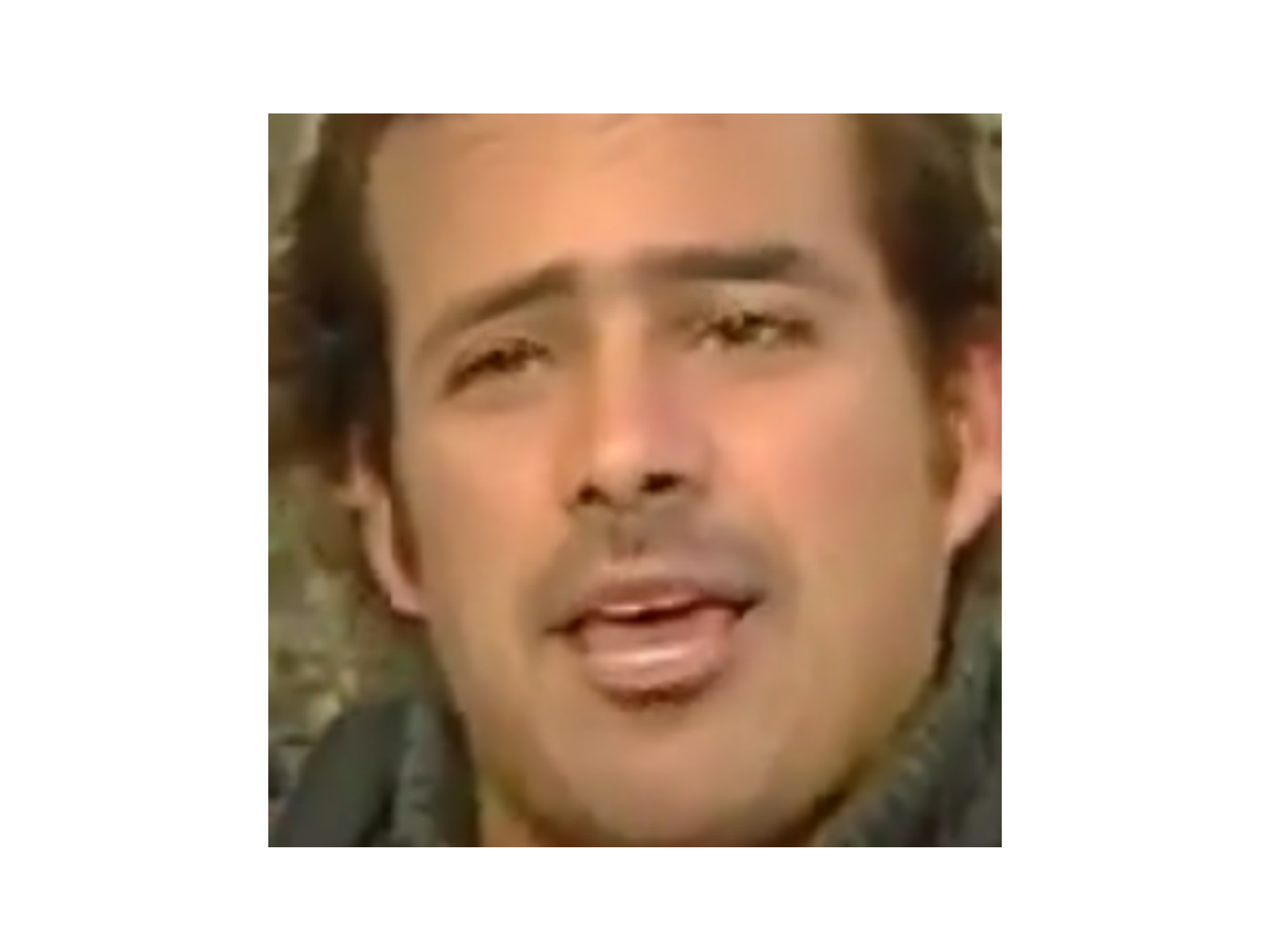}\label{fig:fs_example}}
    \vspace{-1mm}
    \caption{Illustration of (a) a real face, (b) a fake face by face reenactment, and (c) a fake face by face replacement.}
    \label{fig:illustration}
\vspace{-1mm}
\end{figure}

Based on how the faces are manipulated,  face manipulation methods can be divided into two categories, i.e., \emph{face reenactment} and \emph{face replacement}. Given a real face image of one person (called \emph{target face}), face reenactment  \cite{thies2016face2face,thies2019deferred,nirkin2019fsgan} aims to change some properties of the target face image as those of another person's face image (called \emph{source face}), e.g., the expressions, accessories, illuminations, or  shapes of the face. However,   the identity of the target face is preserved. 
Face replacement \cite{deepfakesimplement,faceswap,zakharov2019few} aims to replace the target face with the source face.   
Figure \ref{fig:illustration} illustrates the difference between  face reenactment and face replacement.  In Figure \ref{fig:f2f_example},  the expression and illumination of the face are modified by a face reenactment method, while the identity of the face is preserved. However, in Figure \ref{fig:fs_example}, the face has been totally replaced by another person's face, indicating the identity of the face has changed.

Like face synthesis, many methods have been proposed for face manipulation, including both computer graphics-based methods and deep learning-based methods. 
Next, we  discuss one computer graphics-based method and one representative deep learning-based method for face reenactment and replacement.

\subsubsection{Face Reenactment}

\begin{itemize}

\item {\bf Face2Face}. Face2Face \cite{thies2016face2face} is a computer graphics-based method.  Face2Face first builds a target 3D model for the target faces and a source 3D model for the source faces based on a set of face images.  
Then, given a pair of target face and source face, the attacker transfers the expressions or other properties of the source face to the target face using the two models. Specifically, the attacker computes the parameters (e.g., the expression) of the source face modelled by the source 3D model and uses the target 3D model to generate a fake face with the same parameters.

\item {\bf NeuralTextures}. NeuralTextures \cite{thies2019deferred} is a deep learning-based method that jointly learns neural textures and a rendering neural network based on a set of target faces. The neural textures are high-dimensional feature representations of the target faces, which containc important information about the identity. The rendering neural network takes the neural textures and a uv-map, which is a 2D representation of 3D face information, as its input and reconstructs the face images from them. Given a target face and a source face, the attacker first generates the uv-map of the source face carrying the desired information (e.g., the expression). Then the attacker feeds the uv-map together with the neural textures of the target face into the rendering neural network to re-render a fake face with the identity of the target face and the desired properties of the source face.

\end{itemize}

\subsubsection{Face Replacement}

\begin{itemize}

\item {\bf FaceSwap}. FaceSwap \cite{faceswap} is a computer graphics-based face replacement method. FaceSwap generates a 3D template model with  facial landmarks (e.g., noses, mouths, eyes) that are detected in the target face images. Using the 3D model, FaceSwap projects the face region in the target face image to the source face image by minimizing the difference between the projected facial landmarks and the real facial landmarks in the target face image.

\item {\bf DeepFakes}. With a little abuse of notations, this method is called DeepFakes \cite{deepfakesimplement}. Note that the letter \emph{F} is capitalized in the method name, while it is not in deepfakes referring to forged media.  
DeepFakes leverages autoencoders to perform  face replacement. Specifically, the attacker trains two autoencoders for the source faces and the target faces, respectively. The two autoencoders share the same encoder but have different decoders. To replace a source face as a target face, an attacker can encode the source face with the shared encoder and decode it with the target face's decoder.

\end{itemize}

\subsection{Detecting Deepfakes}

In the past couple of years, many methods have been proposed to detect deepfakes. A deepfake detection system (illustrated in Figure~\ref{overview}) includes two key components, i.e., \emph{face extractor} and \emph{face classifier}. Most deepfake detection systems adopt off-the-shelf face extractor as it is a mature technique while designing customized face classifiers. Roughly speaking, face classifiers can be grouped into two categories, i.e., \emph{heuristics-based} and \emph{neural-network-based}.  Heuristics-based face classifiers \cite{agarwal2019protecting,li2018ictu,matern2019exploiting,yang2019exposing,li2019exposing,frank2020leveraging} leverage some heuristic differences between fake faces and real faces. 
For example, Li et al. \cite{li2019exposing} proposed to capture the face warping artifacts in fake faces. The method is based on the assumption that the quality of the fake faces is lower than that of real faces. Therefore, to match the image quality of the low-resolution fake region and the high-revolution real region, an attacker needs to perform additional face warping, whose artifacts can be used to detect fake faces. 
However, these face classifiers were soon broken by new fake faces. For instance,  deepfake generation methods have been developed to generate high-quality fake faces, breaking the assumption required by \cite{li2019exposing}.

Neural-network-based face classifiers \cite{zhou2017two,afchar2018mesonet,roessler2019faceforensicspp,nguyen2019multi,nguyen2019use,cozzolino2017recasting,bayar2016deep,rahmouni2017distinguishing,wang2019cnngenerated,frank2020leveraging} train  binary neural network classifiers to detect fake faces. Specifically, given a training dataset including both real faces and fake faces, a neural network classifier is trained.  Then, given a new face, 
the classifier  predicts a label ``real'' or ``fake'' for it, indicating whether it is a real one or a fake one. While any neural network classifier could be used, state-of-the-art face classifiers \cite{roessler2019faceforensicspp} leverage the Xception neural network architecture \cite{chollet2017xception} that is pretrained on the ImageNet dataset. Specifically, they fine tune the pretrained Xception neural network using the training  faces as a face classifier.

\subsection{Security of Deepfake Detection} 
The adversarial machine learning community showed that classifiers are vulnerable to adversarial examples \cite{szegedy2013intriguing,carlini2017towards}. Since  a face classifier is a  classifier, it may  also  be vulnerable to adversarial examples. Indeed, several recent studies showed so \cite{gandhi2020adversarial,carlini2020evading,hussain2020adversarial,fernandes2020adversarial}. For instance, Gandhi et al. \cite{gandhi2020adversarial} showed that, via adding a small carefully crafted perturbation to a fake face, a face classifier misclassifies it to a real face, where the fake face with perturbation is known as an adversarial example. 
Carlini et al. \cite{carlini2020evading} proposed several new attacks to generate adversarial examples against  state-of-the-art face classifiers. 
Hussain et al. \cite{hussain2020adversarial} considered real-world adversarial examples that are robust to image and video compression codecs. Fernandes et al. \cite{fernandes2020adversarial} leveraged reinforcement learning to generate adversarial examples against face classifiers. 

However, existing studies only focused on the security of the face classifier against adversarial examples, leaving the security of face extractor and the security of face classifier against cross-method generalization and data poisoning attacks unexplored.

\begin{table*}[!t]\renewcommand{\arraystretch}{1.2}
	\caption{Dataset statistics and  performance of the face classifier trained for each dataset.}
	\vspace{2mm}
	\scalebox{0.77}{\begin{tabular}{|c|c|c|c|c|c|c|c|c|c|c|c|}
		\hline
		\multirow{2}{*}{\small Dataset} & \multirow{2}{*}{\small Dataset source} & {\small Deepfake generation} & \multicolumn{2}{c|}{\small \#training faces} & \multicolumn{2}{c|}{\small \#validation faces} & \multicolumn{2}{c|}{\small \#testing faces} & \multicolumn{3}{c|}{\small Detection performance} \\
		\cline{4-12}
		{} & {} & { method} & {\small Real} & {\small Fake} & {\small Real} & {\small Fake} & {\small Real} & {\small Fake} & {\small Accuracy} & {\small TPR} & {\small TNR}\\
		\hline
		{\small F2F} & {\small FaceForensics++~\cite{roessler2019faceforensicspp}} & {\small Face2Face~\cite{thies2016face2face}} & {\small 367k} & {\small 367k} & {\small 68k} & {\small 68k} & {\small 74k} & {\small 74k} & {\small 0.98} & {\small 0.98} & {\small 0.99}\\
		\hline
		{\small NT} & {\small FaceForensics++~\cite{roessler2019faceforensicspp}} & {\small NeuralTextures~\cite{thies2019deferred}} & {\small 367k} & {\small 292k} & {\small 68k} & {\small 55k} & {\small 74k} & {\small 60k} & {\small 0.94} & {\small 0.90} & {\small 0.97}\\
		\hline
		{\small FS} & {\small FaceForensics++~\cite{roessler2019faceforensicspp}} & {\small FaceSwap~\cite{faceswap}} & {\small 367k} & {\small 291k} & {\small 68k} & {\small 55k} & {\small 74k} & {\small 60k} & {\small 0.99} & {\small 0.98} & {\small 0.99}\\
		\hline
		{\small DF} & {\small FaceForensics++~\cite{roessler2019faceforensicspp}} & {\small DeepFakes~\cite{deepfakesimplement}} & {\small 367k} & {\small 367k} & {\small 68k} & {\small 68k} & {\small 74k} & {\small 73k} & {\small 0.99} & {\small 0.99} & {\small 0.99}\\
		\hline
		{\small DFDC} & \makecell{\small DFDC~\cite{DFDC2020}} & {\small 8 methods \footnotemark} & {\small 362k} & {\small 352k} & {\small 71k} & {\small 68k} & {\small 71k} & {\small 69k}  & {\small 0.98} & {\small 0.99} & {\small 0.98}\\
		\hline
		{\small ALL} & \makecell{\small FaceForensics++~\cite{roessler2019faceforensicspp}\\ \& \small DFDC~\cite{DFDC2020}} & {\small All methods above} & {\small 472k} & {\small 461k} & {\small 89k} & {\small 87k} & {\small 94k} & {\small 92k} & {\small 0.96} & {\small 0.97} & {\small 0.95}\\
		\hline

	\end{tabular}}%
	\label{tab:datasets}
\end{table*}

\section{Measurement Setup}

\footnotetext{DF-128, DF-256, MM/NN, NTH, FSGAN, etc..}
\subsection{Datasets}

We use six datasets from two public large-scale data sources in our experiments, i.e., \emph{F2F}, \emph{NT}, \emph{FS}, \emph{DF}, \emph{DFDC}, and \emph{ALL}. We summarize the statistics of the six datasets in Table~\ref{tab:datasets}.

\myparatight{F2F, NT, FS, and DF} These datasets are extracted from the FaceForensics++ dataset \cite{roessler2019faceforensicspp}. The FaceForensics++ dataset consists of 1,000 real videos from Youtube. Four deepfake generation methods, i.e.,  Face2Face~\cite{thies2016face2face}, NeuralTextures~\cite{thies2019deferred}, FaceSwap~\cite{faceswap}, and DeepFakes~\cite{deepfakesimplement}, are used to manipulate faces in the real videos, which results in 4,000 fake videos in total. 
The videos are compressed using H.264 codec and different video qualities are available. We consider the high quality version of the videos, which are compressed with a constant rate quantization parameter 23. 
We extract the face region in each frame of the videos using the publicly available package Dlib~\cite{dlib09}, and enlarge the located face regions around the center by a factor of 1.3, following the FaceForensics++  paper \cite{roessler2019faceforensicspp}. Moreover, we extract the enlarged face regions from the video frames as face images and resize them to $299\times299$ pixels. The pixel values in the face images are then normalized to [-1,1].

We name the face image dataset consisting of both real faces and fake faces generated by a specific deepfake generation method as the abbreviation of the method. In particular, F2F (NT, FS, or DF) refers to the real faces and the fake faces that are generated by Face2Face (NeuralTextures, FaceSwap, or DeepFakes). 
For each dataset, we split it to a training set, validation set, and testing set following the FaceForensics++ paper \cite{roessler2019faceforensicspp}. Specifically, 720 real videos and their manipulated versions are treated as the training set, 140 real videos and their manipulated versions are treated as the validation set, while the remaining 140 real videos and their manipulated versions are treated as the testing set. In our datasets, the face images successfully extracted from the training/validation/testing videos form the training/validation/testing faces.

\myparatight{DFDC} We extracted this dataset  from the Facebook Deepfake Detection Challenge dataset~\cite{DFDC2020}, which consists of videos  from 3,426 paid actors. The released dataset  contains 19,154 ten-second real videos as well as 100,000 fake videos generated by 8 deepfake generation methods including DFAE~\cite{DFDC2020}, MM/NN~\cite{huang2012facial}, NTH~\cite{zakharov2019few}, and FSGAN~\cite{nirkin2019fsgan}. Moreover, some randomly selected videos are post processed to make the fake videos more realistic, e.g., applying a sharpening filter on the blended faces to increase the perceptual quality of the faces. However, it is unknown which method was used to generate each individual fake video. Therefore, 
unlike the FaceForensics++ dataset, we do not split the dataset based on the deepfake generation methods and treat it as a whole instead. We use 72\% of the  videos as training videos,  14\% as validation videos, and the rest 14\% as testing videos. We extract the face images from the frames of a video following the same  process as the FaceForensics++ dataset. We extract one face image per 50 frames for fake videos and one face image per 10 frames for real videos, considering the different lengths of fake and real videos. We use the face images successfully extracted from the training/validation/testing videos as the training/validation/testing face images in DFDC. Like the FaceForensics++ dataset, we 
resize the face images to $299\times299$ and normalize the pixel values to [-1,1].

\myparatight{ALL}
The ALL dataset is a mix-up of the five face datasets above. Specifically, we randomly select 25\% of the face images in F2F, NT, FS, DF, and DFDC to form the ALL dataset.

\subsection{Training Face Classifiers} \label{sec:detector}
As state-of-the-art face classifiers use the Xception neural network \cite{chollet2017xception}, we train an Xception neural network classifier for each dataset. Specifically, the Xception neural network
 was originally designed for image classification on ImageNet and was pretrained on ImageNet. The last layer is a  fully connected layer with 1,000 neurons. Since deepfake detection is a binary classification problem, the last layer of the pretrained Xception neural network is replaced as a fully connected layer with 2 neurons. Moreover, the parameters for the last layer are randomly initialized.  
 We follow \cite{roessler2019faceforensicspp} to first train the new fully-connected layer for 3 epochs with other layers fixed, and then train the entire network for 15 more epochs. We evaluate the validation accuracy of the model after each epoch and the model with the highest validation accuracy is used as the detector. We use an Adam optimizer with a learning rate $2\times 10^{-4}$ to train the model, which is the same as in \cite{roessler2019faceforensicspp}. 
We train one Xception neural network for each of the six face image datasets, which results in six  face classifiers. 

\subsection{Evaluation Metrics}\label{sec:metrics}
We consider three evaluation metrics to measure the effectiveness of the face classifiers. Specifically, we consider  testing accuracy,  true positive rate, and true negative rate  
as our metrics. When describing true positive rate and true negative rate, we view fake faces as ``positive'' and real faces as ``negative''.  

\myparatight{Accuracy} The accuracy of a face classifier is defined as the fraction of the testing face images that are correctly classified by the face classifier. Accuracy is an aggregated metric, which does not distinguish the detection performance of the real faces and fake faces. Therefore, we further consider true positive rate and true negative rate.

\myparatight{True positive rate} The true positive rate of a face classifier is defined as the fraction of fake face images that are correctly classified as fake by the face classifier. When an attacker tries to evade detection, its goal is to downgrade the true positive rate. 

\myparatight{True negative rate} The true negative rate of a face classifier is defined as the fraction of real face images that are correctly classified as real by the face classifier. True negative rate represents a face classifier's ability to recognize  real faces.

Table~\ref{tab:datasets} shows the performance of the face classifier for each dataset. We observe that all face classifiers are highly accurate. In particular, they achieve accuracies ranging from 0.94 to 0.99. Note that the performance of our detector for DFDC dataset is higher than those of the winning teams in the Facebook competition because they were evaluated on the private testing dataset, which contains unknown post processing and we do not have access to.

\section{Security of Face Extractor} 
\subsection{Experimental Setup}

The security of a deepfake detection system relies on the security of both face extractor and face classifier, e.g., if the face region cannot be extracted accurately from an image,  deepfake detection fails. In this section, we measure the security of face extractor. 
A face extractor aims to extract the face region in an image. When the face extractor cannot find an appropriate face region in an image, it outputs NULL. Note that even if the face extractor does not output NULL, its extracted region may not be the correct face region. We consider the open-source face extractor Dlib~\cite{dlib09}, which was used  by previous work on deepfake detection~\cite{roessler2019faceforensicspp}.  

Recall that  each of our datasets includes image frames obtained from real or deepfake videos. 
We add random Gaussian noise with mean 0 and standard deviation $\sigma$ to each of the three RGB dimensions of each pixel in an image frame. Then, we use Dlib to extract the face region in the noisy image frame. We repeat this experiment for each image frame in a dataset.  Moreover, for a dataset, we define the \emph{face extraction success rate} as the fraction of the image frames in the dataset for which Dlib does not output NULL. Note that our way of defining success rate gives advantages to Dlib, because face extraction may also fail even if Dlib does not output NULL.

\begin{figure}[!t]
    \center
    %\vspace{-3mm}
    \includegraphics[width=0.5\textwidth]{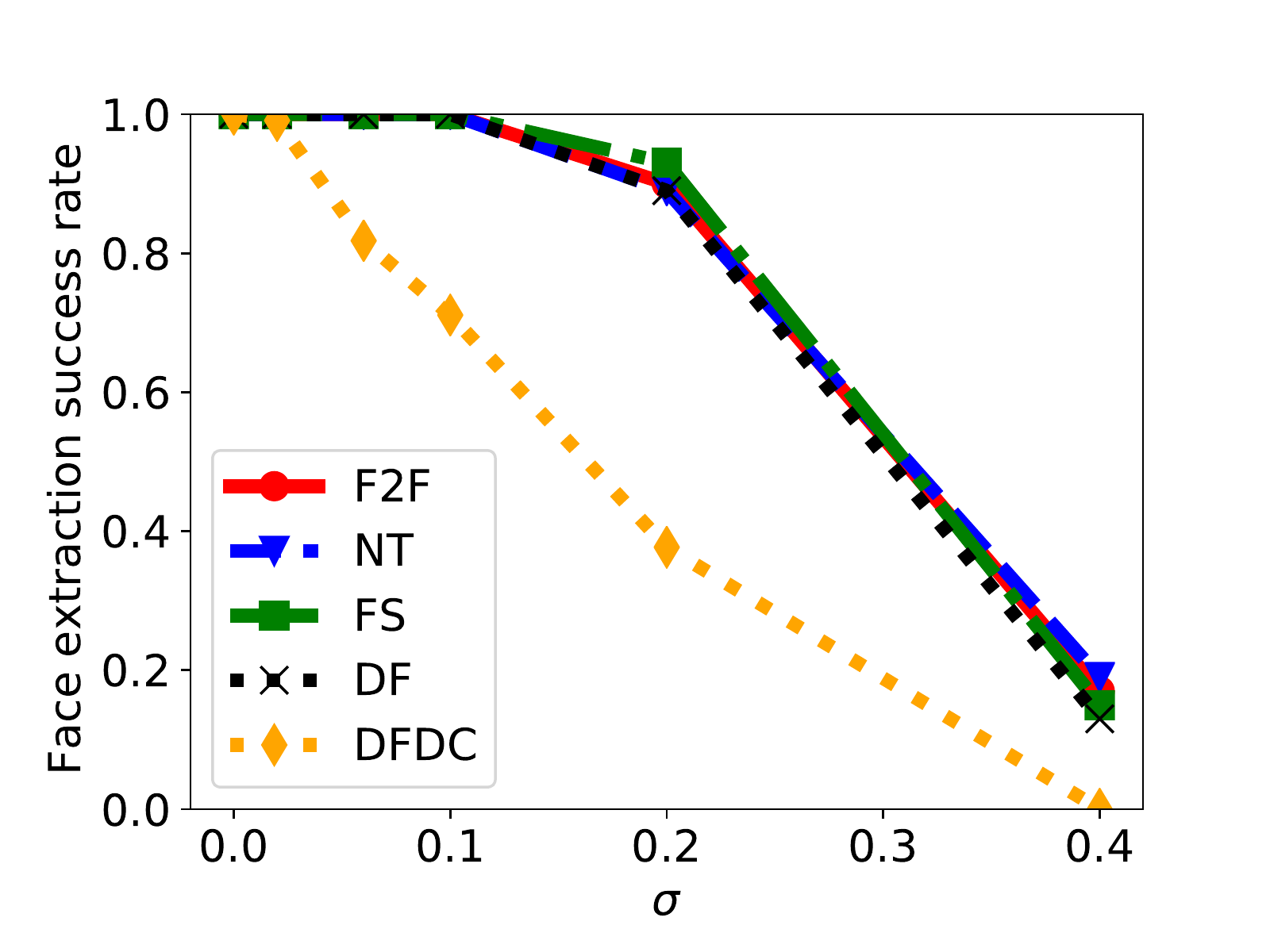}
    %\vspace{-3mm}
    \caption{Face extraction success rate vs. standard deviation of the Gaussian noise added to the image frames.}
    \label{fig:fe}
%\vspace{-5mm}
\end{figure}

\begin{figure}[!t]
    \center
    \includegraphics[width=0.9\textwidth]{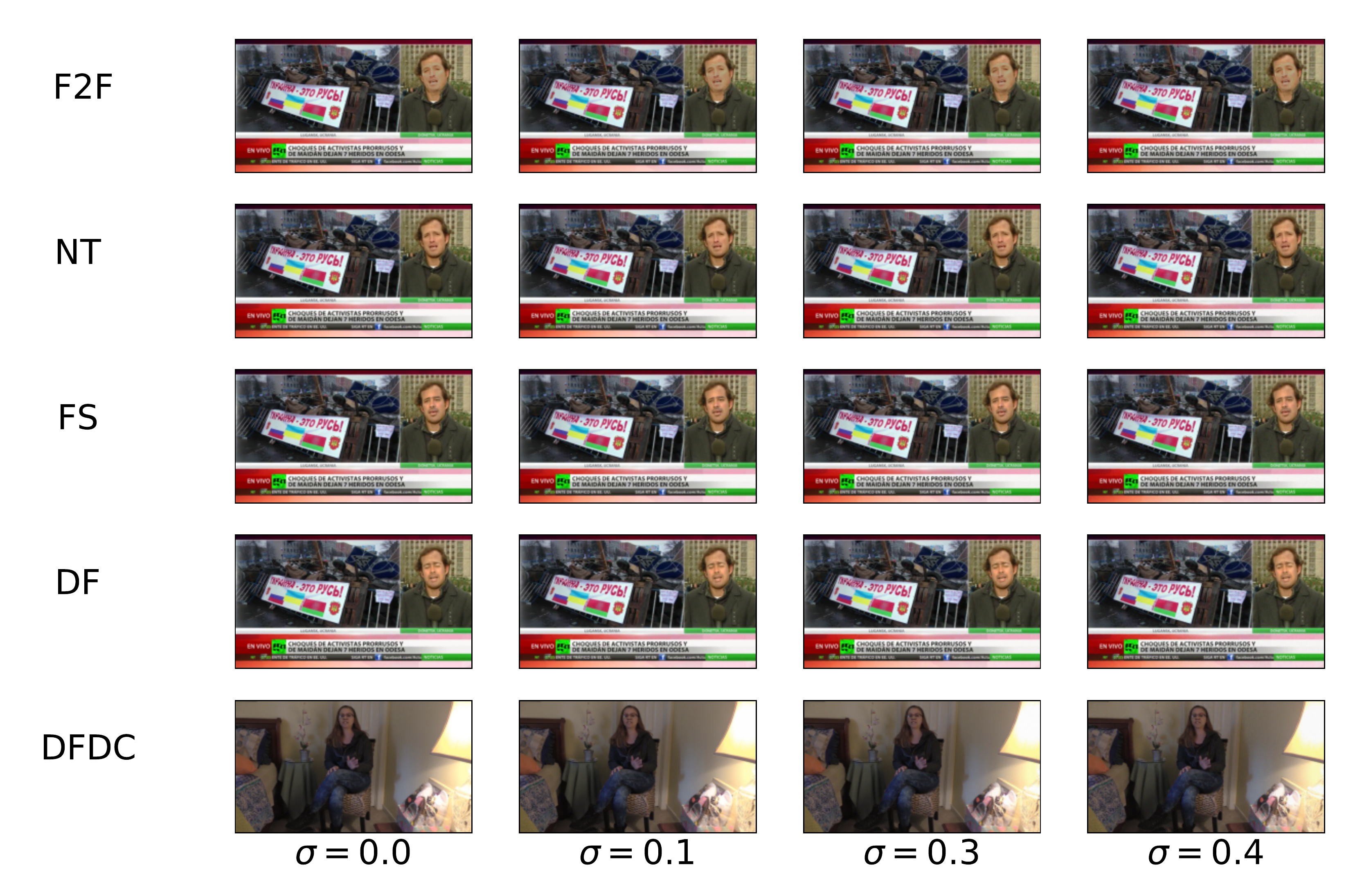}
   \vspace{-3mm}
    \caption{First column:  four deepfake image frames respectively in F2F, NT, FS, and DF generated from the same real image frame and one deepfake image frame in DFDC. Other columns: noisy image frames with different amounts of Gaussian noise.}
    \label{fig:noise_example}
\end{figure}

\subsection{Experimental Results}

Figure \ref{fig:fe} shows the success rates of face extraction for each dataset except ALL when different amounts of Gaussian noise is added to the image frames. We do not show the results for ALL as it is a combination of the other five datasets. We observe that when the noise level $\sigma$ is very small, the Dlib face extractor can extract the faces in most image frames. However, when $\sigma$ increases, the success rate drops quickly, indicating that the face extractor fails to extract the faces when a relatively large Gaussian noise is added to the image frames. 

For instance, when the standard deviation $\sigma$ of the Gaussian noise is 0.2, Dlib outputs NULL for 62\% of image frames in the dataset DFDC and outputs NULL for about 10\% of image frames in the other four datasets; when the standard deviation $\sigma$ of the Gaussian noise increases to 0.4, Dlib outputs NULL for nearly 100\% of image frames in the dataset DFDC and outputs NULL for around 85\% of image frames in the other four datasets. Figure \ref{fig:noise_example}  shows some examples of image frames with different amounts of Gaussian  noise.  
We observe that human can hardly notice the noise even if $\sigma$ is as large as 0.4. In particular, an image frame and its noisy version look the same to human eyes. 

We also observe that the face extraction success rate for DFDC drops faster than those for the other four datasets F2F, NT, FS, and DF. We suspect the reason is that the videos in the FaceForensics++ dataset, which is the source of F2F, NT, FS, and DF, were  selected such that the faces in them  can be extracted easily \cite{roessler2019faceforensicspp}.

\myparatight{Security implications} Our results imply that an attacker can evade detection of its deepfakes via simply adding random Gaussian noise to them to evade a face extractor.

\section{Cross-Method Generalization of Face Classifier}

New deepfake generation methods are continuously developed. Therefore, we are interested in understanding whether a face classifier trained on deepfakes generated by one method can detect deepfakes generated by another method. 

\subsection{Experimental Setup}
Recall that each of the four datasets F2F, NT, DF, and FS include fake faces generated by a particular method. DFDC includes fake faces generated by  8 methods, which include some of the four methods.  ALL is a combination of all the fake faces. Therefore, we use  F2F, NT, DF, and FS to measure cross-method generalization in our experiments as they use different deepfake generation methods. Specifically, we train a face classifier on each of the four datasets as we described in Section~\ref{sec:detector}. Then, we evaluate each face classifier on the testing face images in each dataset.

\begin{table*}[!tb]\renewcommand{\arraystretch}{1.2}
\centering
\caption{The accuracy, true positive rate, and true negative rate of each face classifier on the testing face images in each dataset. Each row represents a face classifier trained on a dataset and each column represents a dataset whose testing face images are used for evaluating face classifiers.}
\subfloat[][Accuracy]{
	\label{tab:unknown_acc}
	\begin{tabular}{|c|c|c|c|c|c|}\hline 
		{\small } & {\small F2F} & {\small NT} & {\small DF} & {\small FS}\\\hline
		{\small F2F} & {\small 0.98} & {\small 0.56} & {\small 0.56} & {\small 0.53}\\\hline
		{\small NT} & {\small 0.52} & {\small 0.94} & {\small 0.54} & {\small 0.61}\\\hline
		{\small DF} & {\small 0.51} & {\small 0.57} & {\small 0.99} & {\small 0.55}\\\hline
		{\small FS} & {\small 0.51} & {\small 0.55} & {\small 0.51} & {\small 0.99}\\\hline
	\end{tabular}
}
\hspace{6mm}
\subfloat[][True positive rate]{
	\label{tab:unknown_tpr}
	\begin{tabular}{|c|c|c|c|c|}\hline 
		{\small } & {\small F2F} & {\small NT} & {\small DF} & {\small FS}\\\hline
		{\small F2F} & {\small 0.98} & {\small 0.02} & {\small 0.03} & {\small 0.06}\\\hline
		{\small NT} & {\small 0.07} & {\small 0.90} & {\small 0.01} & {\small 0.24}\\\hline
		{\small DF} & {\small 0.02} & {\small 0.05} & {\small 0.99} & {\small 0.00}\\\hline
		{\small FS} & {\small 0.02} & {\small 0.01} & {\small 0.02} & {\small 0.98}\\\hline
	\end{tabular}
}
\hspace{6mm}
\subfloat[][True negative rate]{
	\label{tab:unknown_tnr}
    \begin{tabular}{|c|c|c|c|c|}\hline 
		{\small } & {\small F2F} & {\small NT} & {\small DF} & {\small FS}\\\hline
		{\small F2F} & {\small 0.99} & {\small 0.99} & {\small 0.99} & {\small 0.99}\\\hline
		{\small NT} & {\small 0.97} & {\small 0.97} & {\small 0.97} & {\small 0.97}\\\hline
		{\small DF} & {\small 0.99} & {\small 0.99} & {\small 0.99} & {\small 0.99}\\\hline
		{\small FS} & {\small 0.99} & {\small 0.99} & {\small 0.99} & {\small 0.99}\\\hline
	\end{tabular}
}

\label{tab:unknown}
\end{table*}

\subsection{Experimental Results}
Table \ref{tab:unknown} shows the accuracy, true positive rate, and true negative rate of each face classifier on the testing face images in each of the four datasets. We observe the diagonal values in the tables are large. This means that a face classifier trained on  deepfakes generated by some method can accurately detect the deepfakes generated by the same method. However, the off-diagonal accuracies are much smaller, e.g., close to 0.5 (random guessing) in many cases. This means that a face classifier trained on deepfakes generated by some method cannot detect deepfakes generated by other methods. 
We note that the off-diagonal true positive rates are  close to 0 in most cases, while the off-diagonal true negative rates are all close to 1, which means that a face classifier classifies almost all testing face images in a different dataset as real. We suspect the reason is that these four datasets share the same real face images. 

We also train face classifiers using fake faces generated by multiple deepfake generation methods, e.g., F2F + NT, F2F + NT + DF.  
Table~\ref{tab:unknown1} shows the accuracy, true positive rate, and true negative rate of such face classifiers on the testing face images in FS. We observe that even if  a face classifier is trained using fake faces generated by multiple deepfake generation methods, the face classifier still cannot detect fake faces generated by a different method. Note that we did not further include the fake faces in DFDC to train face classifiers, because DFDC may include fake faces that are generated by FS and DFDC does not include information for us to know  which method generated a particular fake face.

\myparatight{Security implications} Our results imply that an attacker can evade detection via generating deepfakes using a new deepfake generation method.

\begin{table*}[!tb]\renewcommand{\arraystretch}{1.2}
\centering
\caption{ The accuracy, true positive rate, and true negative rate of face classifiers on the testing face images in FS. Each row represents a face classifier trained on one or multiple datasets.}
\subfloat[][Accuracy]{
	\label{tab:unknown_acc1}
	\begin{tabular}{|c|c|}\hline 
		{\small }  & {\small FS}\\\hline
		{\small F2F} & {\small 0.56} \\\hline
		{\small F2F + NT} & {\small 0.55} \\\hline
		{\small F2F + NT + DF} & {\small 0.54} \\\hline
	\end{tabular}
}
\hspace{6mm}
\subfloat[][True positive rate]{
	\label{tab:unknown_tpr1}
	\begin{tabular}{|c|c|}\hline 
		{\small }  & {\small FS}\\\hline
		{\small F2F} & {\small 0.03} \\\hline
		{\small F2F + NT} & {\small 0.03} \\\hline
		{\small F2F + NT + DF} & {\small 0.02} \\\hline
	\end{tabular}
}
\hspace{6mm}
\subfloat[][True negative rate]{
	\label{tab:unknown_tnr1}
	\begin{tabular}{|c|c|}\hline 
		{\small }  & {\small FS}\\\hline
		{\small F2F} & {\small 0.99} \\\hline
		{\small F2F + NT} & {\small 0.95} \\\hline
		{\small F2F + NT + DF} & {\small 0.97} \\\hline
	\end{tabular}
}

\label{tab:unknown1}
\vspace{-2mm}
\end{table*}

\section{Security of Face Classifier against Data Poisoning Attacks}
While adversarial examples attack the testing phase of a classifier, data poisoning attacks aim to attack the training phase by polluting the training data such that a corrupted  classifier is learnt. 
In this section, we measure the security of the face classifier against data poisoning attacks.

\subsection{Experimental Setup}
We consider a simple data poisoning attack called  \emph{label flipping attack} and an advanced attack called  \emph{backdoor attack} \cite{Gu17}. For simplicity, we focus on the ALL dataset.

\myparatight{Label flipping attack} Label flipping attack, as its name suggests, 
flips the labels of some training examples. In our deepfake detection, label flipping attack changes the labels of some training real face images to ``fake'' and the labels of some training fake face images to ``real''. In particular, we flip the labels of a certain fraction of the training face images. 
Then, we train the face classifier for the ALL dataset on the training face images including the ones with flipped labels. We evaluate the accuracy, true positive rate, and true negative rate of the corrupted face classifier on the testing face images. Note that we do not change the testing face images.

\begin{figure*}[!t]
    \center
    \subfloat[Trigger]{\includegraphics[width=0.22\textwidth]{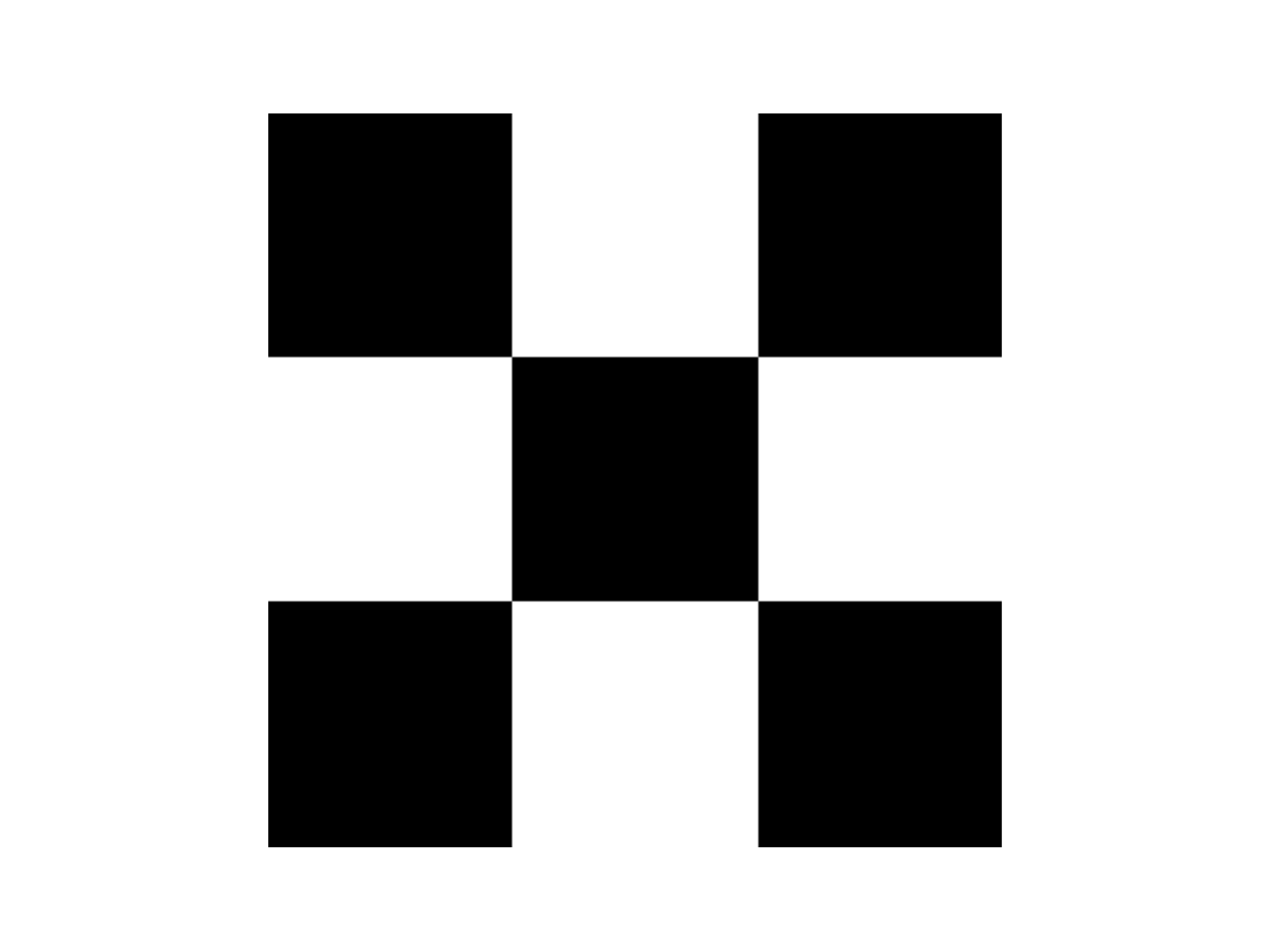}\label{fig:trigger}}
    \subfloat[Face images with trigger embeded]{\includegraphics[width=0.77\textwidth]{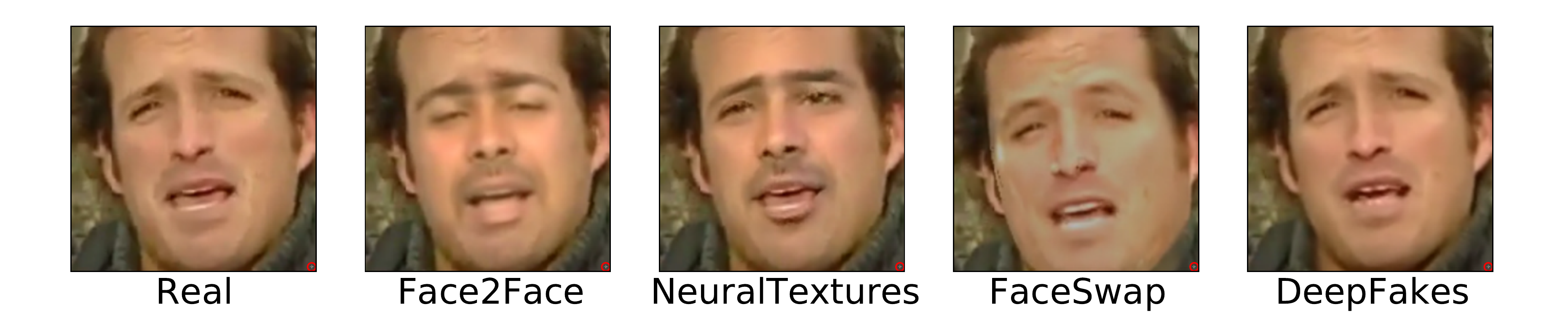}\label{fig:bd_example}}
    \caption{(a) Trigger used in the backdoor attack. (b) A trigger-embedded real face image and four trigger-embedded fake face images generated by four different deepfake generation methods. The trigger is embedded at the bottom right corner of a face image (highlighted by the red circles).}
    \label{fig:bd}
\end{figure*}

\myparatight{Backdoor attack} 
Backdoor attack aims to poison the training examples such that the corrupted classifier predicts an attacker-desired label for any testing example with a \emph{trigger} embedded. In our experiments, we use a chessboard grid as the trigger, which  is shown in Figure \ref{fig:trigger}. Moreover, we set the attacker-desired label as ``real'',  i.e., the corrupted face classifier classifies any face image with the trigger embedded as real. To perform the backdoor attack, we randomly select a fraction of training face images in the ALL dataset. We embed the chessboard grid trigger to the bottom right corner of each of them and set its  label to be ``real''. Figure \ref{fig:bd_example} shows some face images with the trigger embedded. The size of the trigger is  small compared to the size of the face images, i.e., the trigger size is 0.1\% of the image size. Then, we train the face classifier using the training face images including the ones with trigger embedded. We also embed the trigger to each testing face image of the ALL dataset and use them to evaluate the accuracy, true positive rate, and true negative rate of the corrupted face classifier.

\subsection{Experimental Results} 
\begin{figure*}[!t]
    \center
    \subfloat[Label flipping attack]{\includegraphics[width=0.45\textwidth]{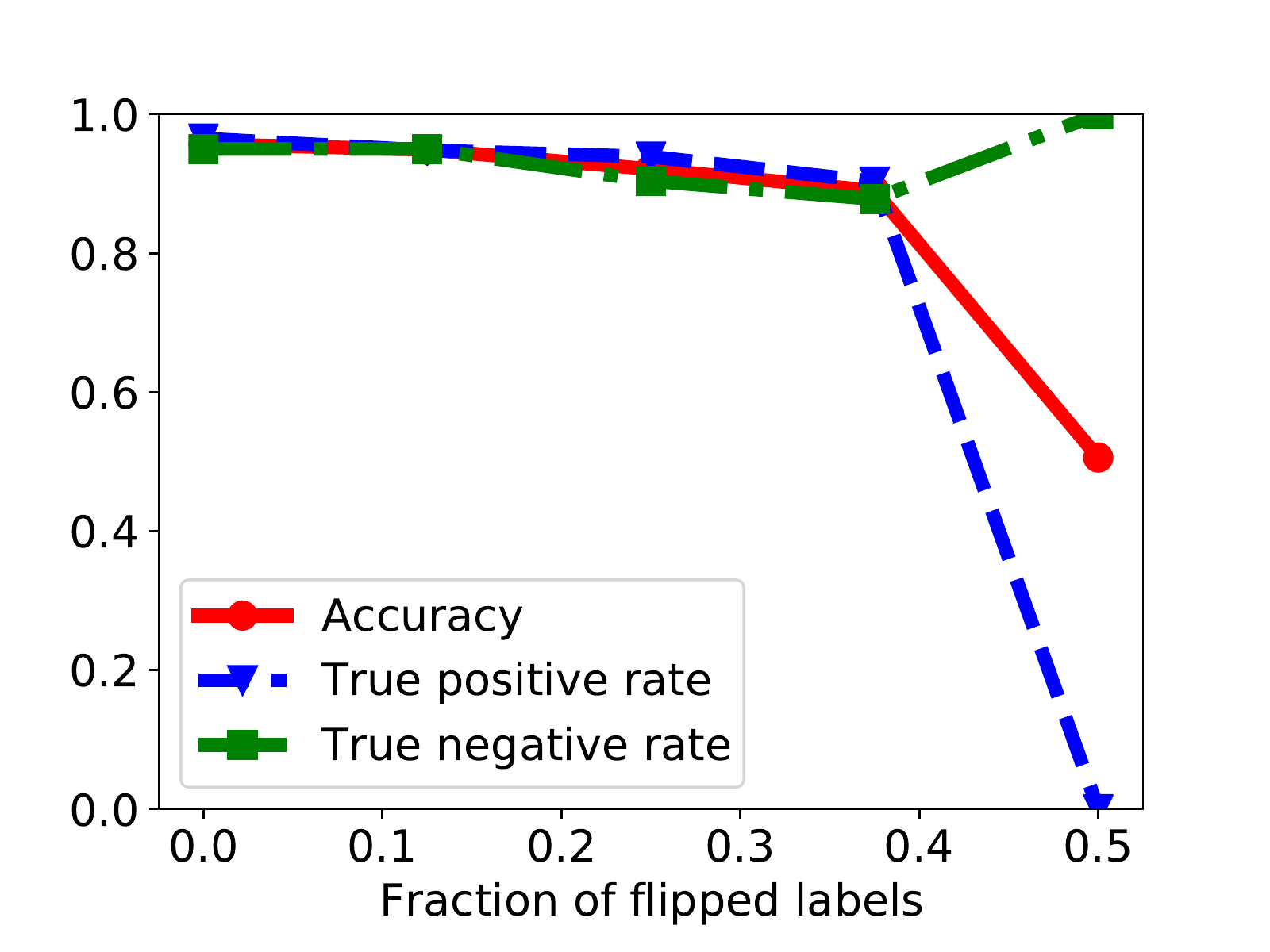}\label{fig:all_lf}}
    \subfloat[Backdoor attack]{\includegraphics[width=0.45\textwidth]{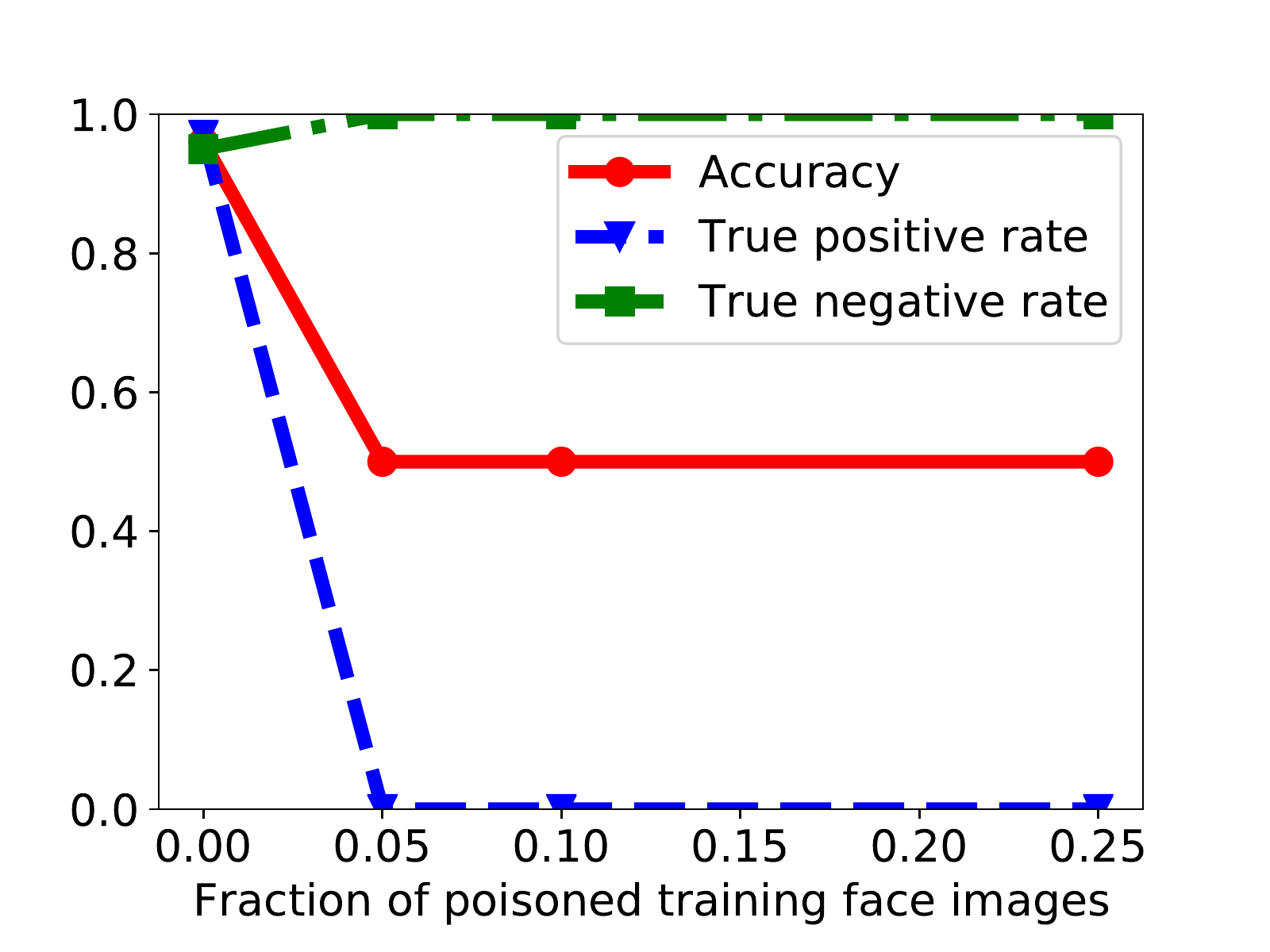}\label{fig:all_bd}}
    \caption{Security of face classifier against label flipping attack and backdoor attack. The ALL dataset is used.} 
    \label{fig:poison}
\end{figure*}

\myparatight{Label flipping attack}
Figure \ref{fig:all_lf} shows the results for label flipping attack. We observe that the face classifier is relatively secure to label flipping attack. Specifically, the accuracy only drops by 0.07 even when the fraction of flipped labels reaches 37.5\%.  We suspect this is because of the redundancy  in the training dataset. As long as the training face images with correct labels are sufficiently more than the training face images with flipped labels, we can learn an accurate face classifier. When a half of the training face images have flipped labels, the learnt face classifier has true positive rate 0 and  true negative rate 1, which indicates that the face classifier classifies all testing face images as real. We suspect this is because 
the ALL dataset has more real face images in the training set and the face classifier learns to predict every image as real. Note that if the fraction of flipped labels  exceeds 0.5, the learnt face classifier is worse than random guessing as more than half of the labels are incorrect.

\myparatight{Backdoor attack} Figure \ref{fig:all_bd} shows the performance of backdoor attack to the face classifier. 
When we do not embed the trigger to any training face images (i.e., the fraction of poisoned training face images is 0), the accuracy, true positive rate, and true negative rate are all close to 1, indicating that 
the face classifier can still correctly classify the testing face images even if we embed the trigger to them. However, when the trigger is embedded into only 5\% of training face images, the true positive rate drops to 0 and the true negative rate becomes 1, indicating that the face classifier classifies all testing face images as real when embedding the trigger to them.

\myparatight{Security implications} Assume a threat model where an attacker can poison some training face images of a face classifier, e.g., flip their labels and embed a trigger to them. For instance, when the training face images are crowdsourced from social media users, an attacker can provide poisoned face images by acting as social media users. Moreover, the attacker can embed the trigger to its fake faces, which is required by backdoor attack. Our measurement results show that an attacker needs to flip the labels of a large fraction (e.g., $>$40\%) of the training face images in order to attack a face classifier via label flipping attack. However, an attacker can evade detection via backdoor attack that only poisons a small fraction (e.g., 5\%) of the training face images. 

\section{Discussion and Limitations}

\myparatight{Leveraging robust face extractor and face classifier} Our measurement results show that face extractor and face classifier are not secure against perturbations (e.g., adversarial examples) added to testing images and backdoor attacks. We note that how to defend against adversarial examples and backdoor attacks is still an open challenge, though the adversarial machine learning community has developed multiple methods \cite{madry2017towards,cao2017mitigating,cohen2019certified,wang2019neural,gao2019strip,liu2019abs} to enhance classifiers' robustness against them. Among the different methods, \emph{adversarial training} \cite{madry2017towards} and \emph{randomized smoothing} \cite{cao2017mitigating,cohen2019certified,jia2019certified} achieve state-of-the-art robustness against adversarial examples, while ensemble methods~\cite{jia2020intrinsic,wang2020certifying} achieve state-of-the-art robustness against backdoor attacks. 

In particular, adversarial training adds adversarial examples of the training examples to augment the training dataset, while randomized smoothing builds a smoothed classifier by  randomizing an input and provides probabilistic certified robustness guarantee  of the smoothed classifier. Randomized smoothing ensures that no adversarial perturbation smaller than a threshold can change the predicted label of a given testing example. However, these methods sacrifice a classifier's accuracy when no perturbation is added to the testing examples~\cite{madry2017towards,cohen2019certified}, i.e., a face classifier built by these methods has a lower accuracy even if an attacker does not add any perturbation to its fake faces. Moreover, these methods can only defend against very small perturbations, i.e., an attacker can still evade detection once adding large enough perturbations to its fake faces.  Ensemble methods train multiple classifiers and take a majority vote among them to predict the label of a testing example. The predicted label is unaffected by a small number of poisoned training examples. However,  a face classifier built by such methods also has a lower accuracy even if an attacker does not perform backdoor attacks. 

Neural cleanse \cite{wang2019neural} was proposed as a defense against backdoor attacks. Specifically, neural cleanse can identify potential backdoors in a classifier and reconstruct the trigger. Neural cleanse is based on an assumption that all testing images embedded with a specific trigger will be predicted as the same target label. Therefore, the trigger can be reverse engineered by searching for the minimum perturbation that can change the classification results of all testing examples to a certain label. Once the trigger is reconstructed, neural cleanse uses input filters, neural pruning, or machine unlearning to eliminate the effect of the backdoor embedded in the classifier. However, neural cleanse cannot detect source-label-specific backdoor attacks, where the backdoor is designed to be effective only for a subset of source testing examples, e.g., face images with blonde hair. In this scenario, the classification results for face images whose hair is not blonde will not be affected by the trigger. Therefore, the assumption that Neural cleanse relies on does not hold and it fails to detect the backdoor \cite{wang2019neural}.

\myparatight{Deepfake video} In this work,  we consider deepfake detection for a static face image. In practice, we may have access to deepfake videos. Therefore, a deepfake detector can consider the statistical information between the image frames in a video to classify it to be real or fake. For instance, one way is to classify each  frame of a video as real or fake, and then take a majority vote among the labels of the frames as the label of the entire video. 
Another intuitive way to deal with videos is to use sequential information of a video. For instance, the detector can track the light source in the video and classify  the video as fake if there are inconsistencies in the light source location \cite{johnson2007exposing}. Audio information in a video may also be used to aid  detection of deepfake videos. However,  an attacker does not need to manipulate the audios and one of the leading teams in the Facebook Deepfake Detection Challenge competition found that audio may not  necessarily be helpful for deepfake detection \cite{audionothelp}.

\myparatight{Leveraging network security solutions} Instead of detecting abnormality in the content of an image or video, we can also block the spread of  deepfakes  from the network security perspective. In particular,  deepfakes are often propagated via social media and they may be propagated by fraudulent users such as fake users and compromised users. Therefore, we can detect fraudulent users in social media who propagate deepfakes and limit the impact of deepfakes. Many approaches have been proposed to detect fraudulent users. These approaches leverage user registration information, user behavior, content generated by  user, and/or social graphs between users~\cite{gong2014sybilbelief,yuan2019detecting,cao2014uncovering,wang2013you,danezis2009sybilinfer,egele2013compa,wang2018graph,wang2017sybilscar,jia2017random,wang2017gang}. Although these methods cannot detect all fraudulent users, they may increase the bar for attackers to maintain them and spread deepfakes. 
\section{Conclusion} 
We evaluated the security of the state-of-the-art deepfake detection methods using six datasets from two large-scale public data sources. Our extensive experiments show that although the detectors can achieve high accuracies in non-adversarial settings, a face extractor is not secure against random Gaussian  noise added to the images. 
Moreover, we found that  a face classifier trained using fake faces generated by some deepfake generation methods cannot detect fake faces generated by a different method; and a face classifier is not secure against backdoor attacks.  Our results highlight that the major challenge of deepfake detection is to enhance its security in adversarial settings.

\section*{Acknowledgements}
We thank the anonymous reviewers for insightful reviews. We also thank Xiaohan Wang for discussion and processing datasets for experiments on cross-method generalization. This work was partially supported by NSF grant No.1937786.

{
\bibliographystyle{splncs04.bst}
\bibliography{refs}
}

\end{document}